\documentclass{aa}  

\usepackage{natbib}
\usepackage{graphicx}
\usepackage{txfonts}
\begin{document}

\title{Simultaneous 183 GHz H$_{2}$O Maser and SiO Observations Towards Evolved Stars Using APEX SEPIA Band 5}

\titlerunning{183 GHz H$_{2}$O Maser and SiO Observations Towards Evolved Stars}

\authorrunning{Humphreys et al.}

   \author{E. M. L. Humphreys
          \inst{1}
          \and
           K. Immer
          \inst{1}
          \and
          M. D. Gray
          \inst{2}
           \and
          E. De Beck
          \inst{3}
          \and
          W. H. T. Vlemmings
          \inst{3}
          \and
         A. Baudry
          \inst{4}
          \and
          A. M. S. Richards
          \inst{2}
          \and
          M. Wittkowski
          \inst{1}
          \and
          K. Torstensson
          \inst{1}
          \and
          C. De Breuck
          \inst{1}
          \and
          P. M{\o}ller
          \inst{1}
          \and
          S. Etoka
          \inst{5}    
          \and
          M. Olberg
          \inst{3}   
          }

   \institute{European Southern Observatory (ESO), 
              Karl-Schwarzschild-Str. 2,
              85748 Garching bei Munchen, Germany \\
              \email{ehumphre@eso.org}
         \and
             Jodrell Bank Centre for Astrophysics, School of Physics and Astronomy, University of Manchester, Manchester M13 9PL, UK
         \and
          Department of Earth and Space Sciences, Chalmers University of Technology, Onsala Space Observatory, S-439 92 Onsala,  Sweden
          \and
          Laboratoire d'astrophysique de Bordeaux, Univ. Bordeaux, CNRS, B18N, all\'ee Geoffroy Saint-Hilaire, F-33615 Pessac, France
          \and
          Hamburger Sternwarte, Universit\"at Hamburg, 21029 Hamburg, Germany        
             }

   \date{Received ; accepted }

 
  \abstract
   {}
   {The aim is to investigate the use of 183 GHz H$_{2}$O masers for characterization
    of the physical conditions and mass loss process in the circumstellar envelopes of evolved stars.}
   {We used APEX SEPIA Band 5 (an ALMA Band 5 receiver on the APEX telescope) to observe the 183 GHz
    H$_{2}$O line towards two Red Supergiant (RSG) and three Asymptotic Giant Branch (AGB) stars. Simultaneously,
    we observed the J=4$-$3 line for $^{28}$SiO v=0, 1, 2 and 3, and for $^{29}$SiO v=0 and 1.
    We compared the results with simulations and radiative transfer models for H$_{2}$O and SiO, and examined
    data for the individual linear orthogonal polarizations.}
   {We detected the 183 GHz H$_{2}$O line towards all the stars with peak flux densities $>$100 Jy, including a new detection from VY CMa. 
   Towards all five targets, the water line had indications of being due to maser emission and had higher peak flux 
   densities than for the SiO lines. The SiO lines appear to originate from both thermal and maser processes.
    Comparison with simulations and models indicate that 183 GHz maser emission is likely to extend to
    greater radii in the circumstellar envelopes than SiO maser emission and to similar or greater radii than water masers at 22, 321
    and 325 GHz. We speculate that a prominent blue-shifted feature in the W Hya 183 GHz  spectrum is 
    amplifying the stellar continuum, and is located at a similar distance from the star as mainline OH maser emission.
    We note that the coupling of an SiO maser model to a hydrodynamical pulsating model of an AGB star
   yields qualitatively similar simulated results to the observations. 
    From a comparison of the individual polarizations, we find that the SiO maser linear polarization fraction of
several features exceeds the maximum fraction allowed under standard
maser assumptions and requires strong anisotropic pumping of the maser
transition and strongly saturated maser emission.  The low polarization fraction of the H$_2$O maser however, fits with the expectation for a non-saturated maser.}
   {183 GHz H$_{2}$O masers can provide strong probes of the mass loss process of evolved stars. Higher angular resolution
   observations of this line using ALMA Band 5 will enable detailed investigation of the emission location in circumstellar envelopes
   and can also provide information on magnetic field strength and structure.}


   \maketitle
%

\section{Introduction}

Understanding the mass loss mechanism of Asymptotic Giant Branch (AGB) and
Red Supergiant (RSG) stars is necessary for  determination of the formation processes of Planetary Nebulae and
 core-collapse Supernovae \citep[e.g.,][]{Herwig2005,Smith2014}. In addition, the substantial mass
loss of cool evolved stars makes significant contributions to dust and molecular 
return to the interstellar medium and thereby the chemical evolution of galaxies \citep[e.g.,][]{Javadi2016}.

\begin{table}
\begin{center}
\caption{Line List}
\label{linelist}
\begin{tabular}{ccc}
\hline
\hline
Frequency &  Transition    & E$_u$/k   \\ 
 (GHz)    &                &  (K)  \\
\hline
\multicolumn{3}{c}{para-H$_{2}$O} \\
183.310   &      3$_{13}$--2$_{20}$& 205  \\
\hline
\multicolumn{3}{c}{$^{28}$SiO} \\
173.688     & v=0 J=4$-$3 &  21    \\ 
172.481     & v=1 J=4$-$3 & 1790    \\ 
171.275     & v=2 J=4$-$3 & 3542    \\ 
170.070     & v=3 J=4$-$3 & 5277    \\
\hline
\multicolumn{3}{c}{$^{29}$SiO}\\
171.512     & v=0 J=4$-$3 &  21   \\
170.328     & v=1 J=4$-$3 &  1779   \\            
\hline 
\end{tabular}
\end{center}
\end{table}

\begin{table*}[tbh]
           \caption{\label{properties} Stellar Sample}
           \begin{centering}          
   \begin{tabular}{lccccc}
   \hline\hline
            Star                   & VY CMa       & VX Sgr        &     W Hya     &  U Her       &  R Aql \\
	                              &                     &                    &                     &                  &         \\  
            \hline
            RA                     &  07:22:58.33   & 18:08:04.05   & 13:49:02.00   & 16:25:47.47  &  19:06:22.25  \\
            Dec                    & $-$25:46:03.2&$-$22:13:26.6  & $-$28:22:03.5 & $+$18:53:32.9&  $+$08:13:48.0 \\
Distance\tablefootmark{a}           &     1.2 kpc  &   1.6 kpc     &  104 pc         &  266 pc        &   422 pc           \\
Type\tablefootmark{b}                  &     RSG      &  RSG         & Mira$/$SRa     &  Mira        &  Mira           \\
Spectral Type\tablefootmark{b}     &       M5      &M4$-$M10       & M7$-$M9     & M6.5$-$M9.5  &   M5e$-$M9e     \\
Current Period (days)\tablefootmark{b}   &    1600        &  732          &  390          &  404         &  270.5       \\
Magnitude Range\tablefootmark{b}  & 6.5$-$9.6 V  &6.5$-$14 V    &5.6$-$9.6 V  & 6.4$-$13.4 V &  5.5$-$12 V  \\
Masers            &SiO,H$_{2}$O,OH&SiO,H$_{2}$O,OH&SiO,H$_{2}$O,OH &SiO,H$_{2}$O,OH&SiO,H$_{2}$O,OH\\
Mass Loss Rates\tablefootmark{c} ($M_{\odot}$yr$^{-1}$)&1.8$\times$ 10$^{-4}$&6.3$\times$10$^{-5}$&1.4$\times$10$^{-7}$&1.6$\times$10$^{-7}$ & 2-14$\times$10$^{-6}$\\
            \hline
       \end{tabular}
\end{centering}
\tablefoot{
\tablefoottext{a}{Distances are from: W Hya and R Aql \citep[Hipparcos,][]{vanLeeuwen2007}, VY CMa \citep{Choi2008,Zhang2012}, VX Sgr \citep{Chen2007} and U Her \citep{Vlemmings2007}}.
\tablefoottext{b}{From the AAVSO, except for the period of VY CMa which is from \citet{Kiss2006}.}
\tablefoottext{c}{Mass loss rates, scaled to the distances used here, are from \citet{deBeck2010} for VY CMa, VX Sgr and W Hya. For U Her  and R Aql they are from \citet{Yates1995} and  \citet{ZhaoGeisler2012} respectively.}}
   \end{table*}
   
     \begin{figure}
   \includegraphics[width=8.0cm,angle=-90]{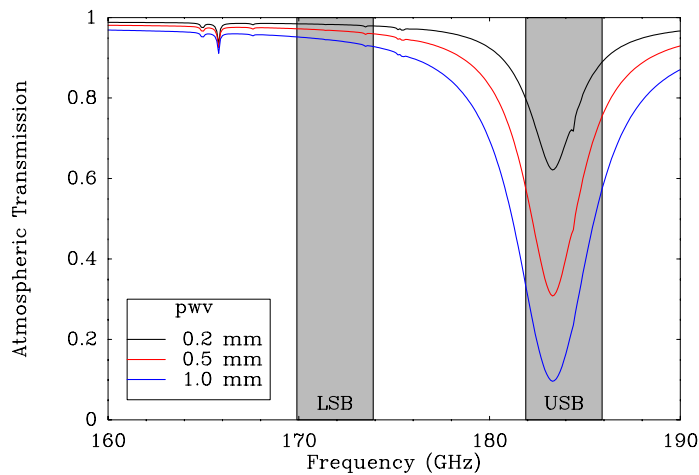}
   \caption{Zenith atmospheric transmission for precipitable water vapour (pwv) values of 0.2, 0.5 and 1 mm at Llano de Chajnantor. Values are plotted from the ATM model \citep{Pardo2001}. 
   Frequency coverage of the observations is shaded in grey for the sidebands. Each sideband consists of 4 GHz continuous frequency coverage. 
   The rest frequency of the 3$_{13}$--2$_{20}$ H$_{2}$O line is 183.310 GHz.
   }              \label{transmission}%
    \end{figure}

   \begin{figure}
   \includegraphics[width=8.0cm,angle=-90]{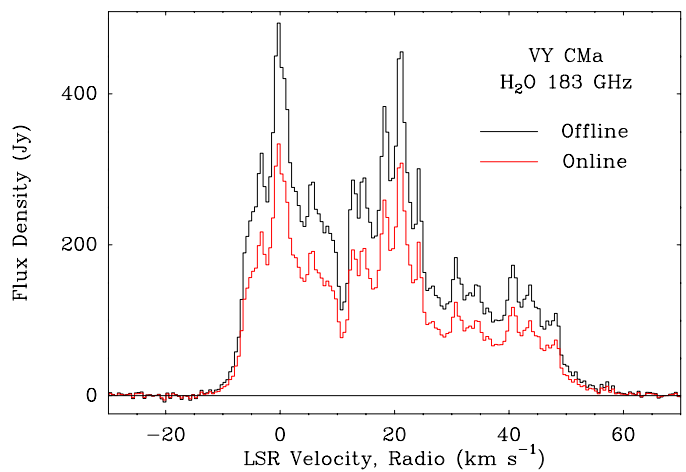}
   \caption{Comparison of the spectra obtained using offline and online opacity calibration for the 183 GHz H$_{2}$O line towards VY CMa, plotted at 0.5 kms$^{-1}$ resolution. 
   Use of the offline opacity correction increases peak flux density of the 183 GHz water line in VY CMa by 48\%, where atmospheric transmission varies significantly across the sub-bands making up the USB. It makes little difference to the SiO lines that are in a region of flatter atmospheric transmission in the LSB, as expected. Throughout this paper values and figures from the offline calibration are used.}
              \label{calib}%
    \end{figure}




\begin{figure*}
\includegraphics[width=9.0cm]{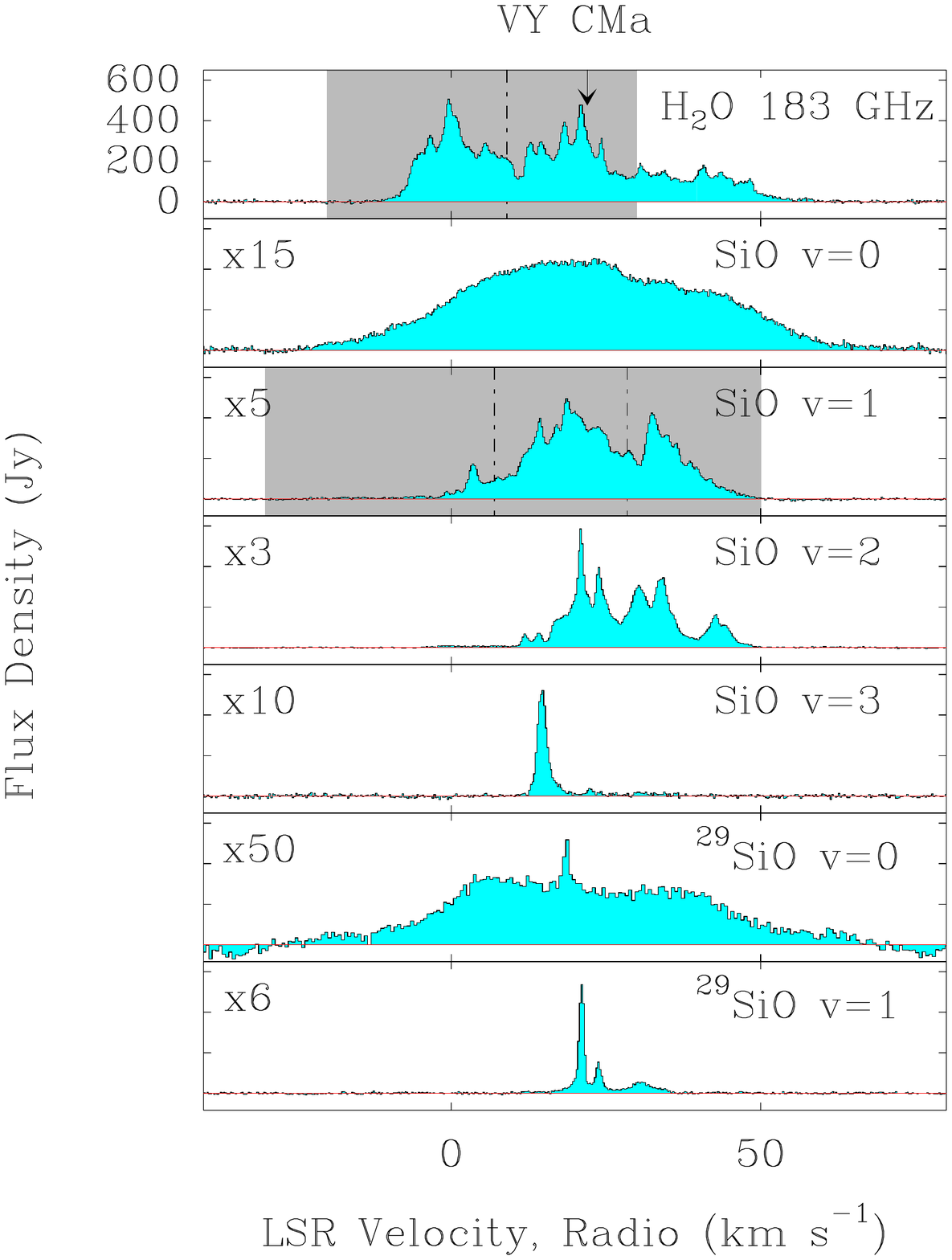}\includegraphics[width=9.0cm]{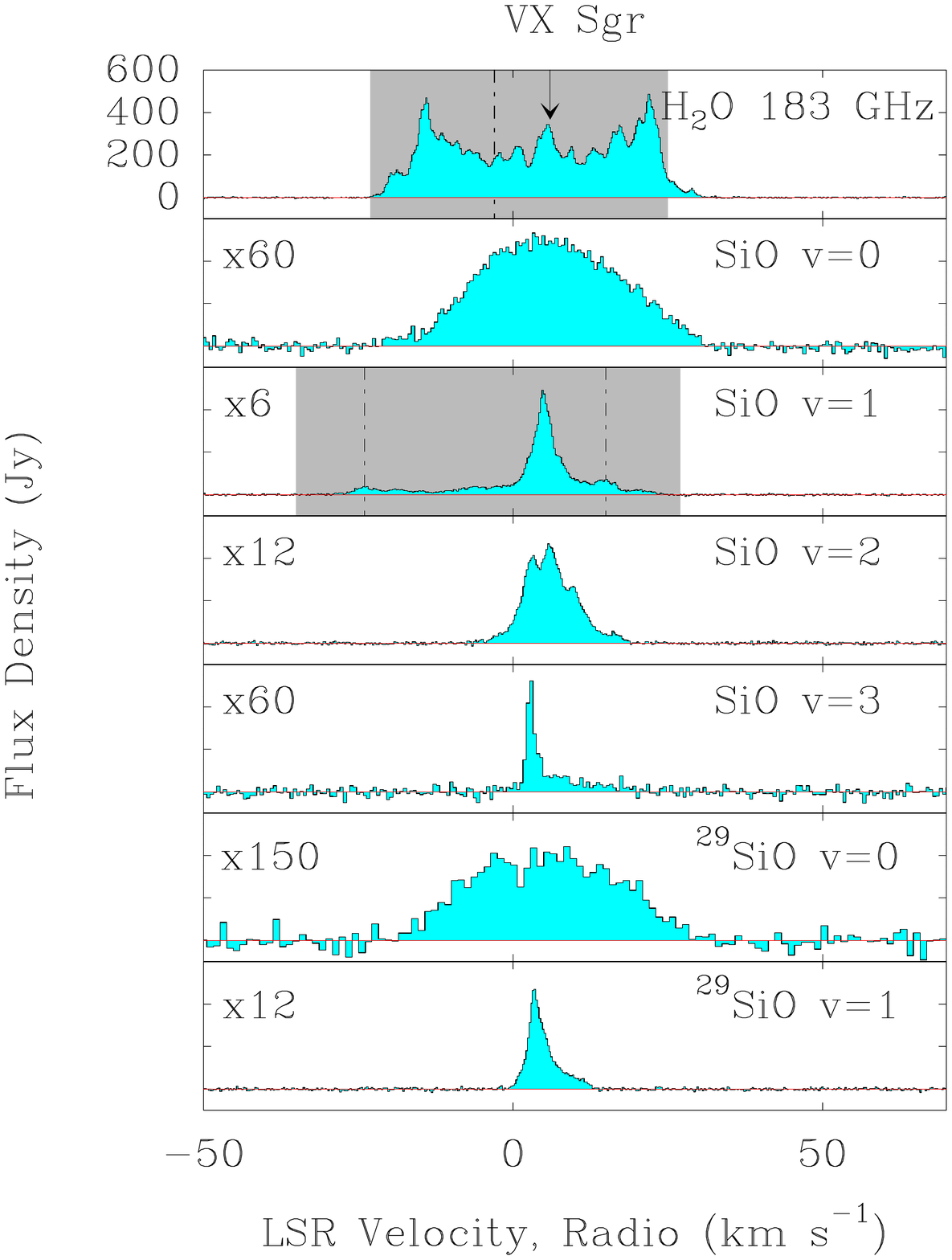}
\caption{VY CMa and VX Sgr observations using APEX SEPIA Band 5. All of the SiO lines originate from the J=4$-$3 transition. The arrow in the upper panel marks the stellar systemic velocity. 
Boxes shaded grey indicate the velocity ranges that can be contaminated by sideband leakage (Sect.~\ref{sideband}), with 
dashed vertical lines indicating the velocity to which peak(s) of emission
from the other sideband would ``ghost'' i.e.,  the dashed lines in the SiO v=1 panels indicate the velocities at which peaks from the H$_{2}$O spectrum could create ghosts. The dashed lines in the
H$_{2}$O panels indicate the velocities at which peaks from the SiO v=1 J=4$-$3 spectrum could create ghosts.}
 \label{spectrum1}%
 \end{figure*}

One set of probes for helping to understand evolved star mass loss
is provided by the circumstellar masers, which are commonly present.
In oxygen-rich evolved stars, SiO masers form typically within several stellar
radii ($\sim$2$-$5 R$_*$). They trace both infall and outflow apparently associated with
stellar pulsation, plus complex motions \citep[e.g.,][]{Assaf2011,Gonadikis2013}. 
22 GHz H$_{2}$O masers and 1665/7 MHz OH masers form at larger radii in the circumstellar envelope
(CSE). They trace acceleration of the wind in the dust formation zone, at approximate distances 10$-$100 R$_*$
\citep[e.g.,][]{Bains2003,Richards2012}.
Finally, 1612 MHz OH masers are observed at even larger radii, in the steadily outflowing wind at several hundred to 1000 R$_{*}$ \citep{Etoka2004,Gray2012}. 
The masers can thus provide information on the physical
conditions for a range of zones in the CSE, as well as tracing dynamics and magnetic field strength and
morphology via the Zeeman Effect \citep{Vlemmings2014}. Masers from SiO and H$_2$O transitions occur throughout the APEX
and ALMA bands \citep[e.g.,][]{Humphreys2007,Gray2012}. \citet{Richards2014} have mapped for the first time
water maser emission at 321, 325 and 658 GHz towards an evolved star, VY CMa, using ALMA.

During Science Verification of the SEPIA\footnotemark[1] Band 5 receiver on APEX (an ALMA Band 5 receiver) \citep[][]{Billade2012,Immer2016}
we observed the 183 GHz H$_2$O line toward a small sample of stars, simultaneously
with a number of SiO J=4$-$3 lines.
\citet[][hereafter GA98]{Gonzalez1998} previously observed the 183 GHz line towards a sample of 23 stars using the
IRAM 30-m. GA98 established that the 183 GHz line is not very variable in comparison with
the 22 GHz H$_{2}$O masers, and can be very strong with line peaks of up to 450 Jy. GA98 also found that
the single-dish line profile of the 183 GHz maser differs as a function of stellar mass loss rate, as has
also been observed for some 22 GHz H$_2$O masers, going from a single-peak at the stellar velocity in stars of lower mass loss rate, to
double-peaked lines that bracket the stellar velocity at high mass loss rate (indicative of going from predominantly 
tangential amplification near to the star to radial maser amplification further from the stars). In this paper we present the APEX SEPIA observational results
and compare them against simulations and models, and make predictions for ALMA Band 5 observations.

\footnotetext[1]{Swedish-ESO PI receiver for APEX}


\section{Observations}

The observations were made during APEX SEPIA Band 5 Science Verification under programme codes
ESO 095.F-9806(A) and Onsala 095.F-9400(A). U Her and W Hya were observed between 21 to 22 May 2015.
VY CMa, VX Sgr and R Aql  were observed between July 13 to 16 2015.  SEPIA Band 5 is a dual polarization sideband-separating
(2SB) receiver (Belitsky et al., in prep). It was tuned to 183.9 GHz at the centre of the upper sideband (USB) such that the frequency coverage of
the observations was 181.9 to 185.9 GHz (USB) and 169.9 to 173.9 GHz in the lower sideband (LSB; Figure~\ref{transmission}). The frequency
coverage of 4 GHz in each sideband is achieved by using two overlapping 2.5 GHz sub-bands. The observations were made in position-switching mode. The half-power
beam width (HPBW) at 183.3 GHz is 31.8\arcsec. The targets were observed for times ranging from
6 to 40 minutes. The precipitable water vapour during the observations varied between 0.25 and 0.64 mm.

Standard calibrations were made at the Observatory. For the observations made in May, the majority
of the data were taken with an internal cold load that leads to a higher receiver temperature by a factor of two than
the observations made in July, which used an external, facility calibration unit. This affected steps in the subsequent calibration. For this 
reason the fluxscale of the observations made towards U Her and W Hya should not be trusted. However
line ratios of the SiO lines can still be used. We estimate the fluxscale is uncertain by up to
a factor of two for the July 2015 observations of VY CMa, VX Sgr and R Aql.

The observatory delivered data with an online  standard 
opacity correction. This uses one atmospheric opacity value across a 2.5 GHz sub-band and
it works well when the opacity is reasonably flat. However, in the case
of observations near to the 183.3 GHz water line it is better to perform an offline recalibration using
opacity values obtained on the resolution of the atmospheric model used \citep[ATM,][]{Pardo2001}. The APEX offline recalibration yields 
opacity values determined in chunks of 128 channels. Comparison between
the spectra obtained using online and offline calibrated data indicates the online calibration underestimates the emission around 183 GHz significantly
 \citep{Immer2016}, 
e.g. for VY CMa by 48 \% (Figure~\ref{calib}). In this paper, we use offline recalibrated data throughout.

\begin{figure*}
\includegraphics[width=9.0cm]{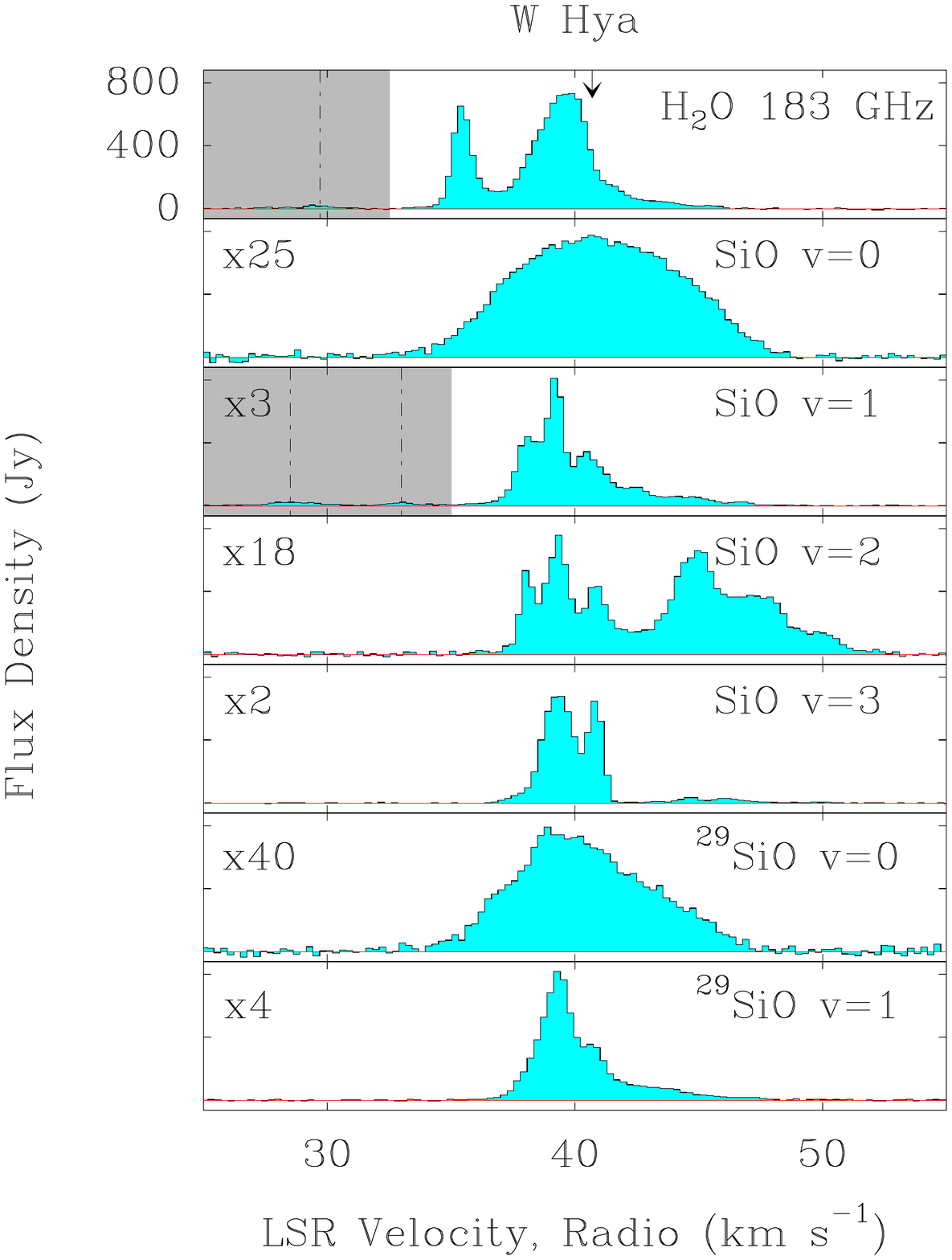}\includegraphics[width=9cm]{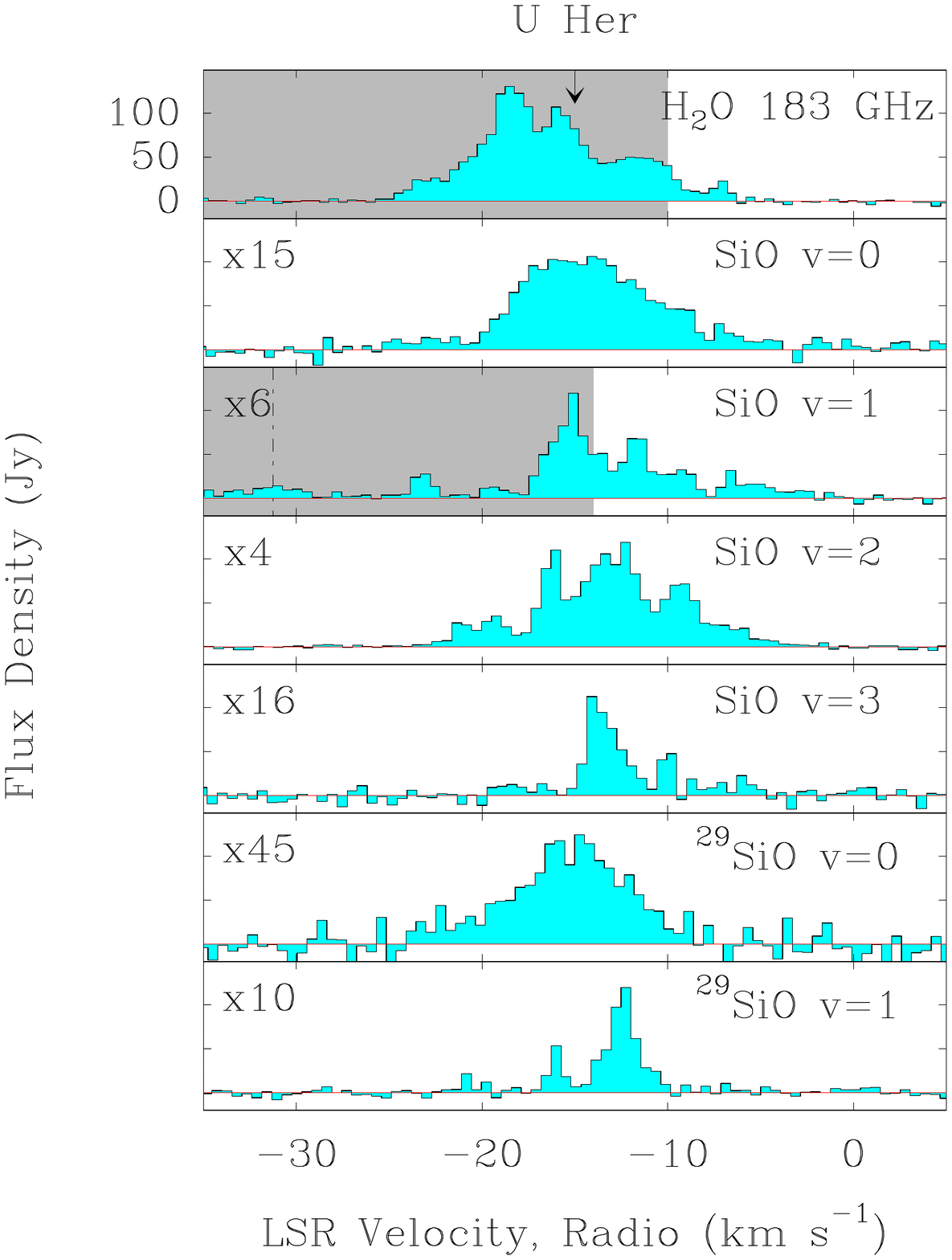}
\caption{W Hya and U Her observations using APEX SEPIA Band 5. See caption of Figure~\ref{spectrum1} for more details.}
\label{spectrum2}%
\end{figure*}

\begin{figure}
   \centering
   \includegraphics[width=9cm]{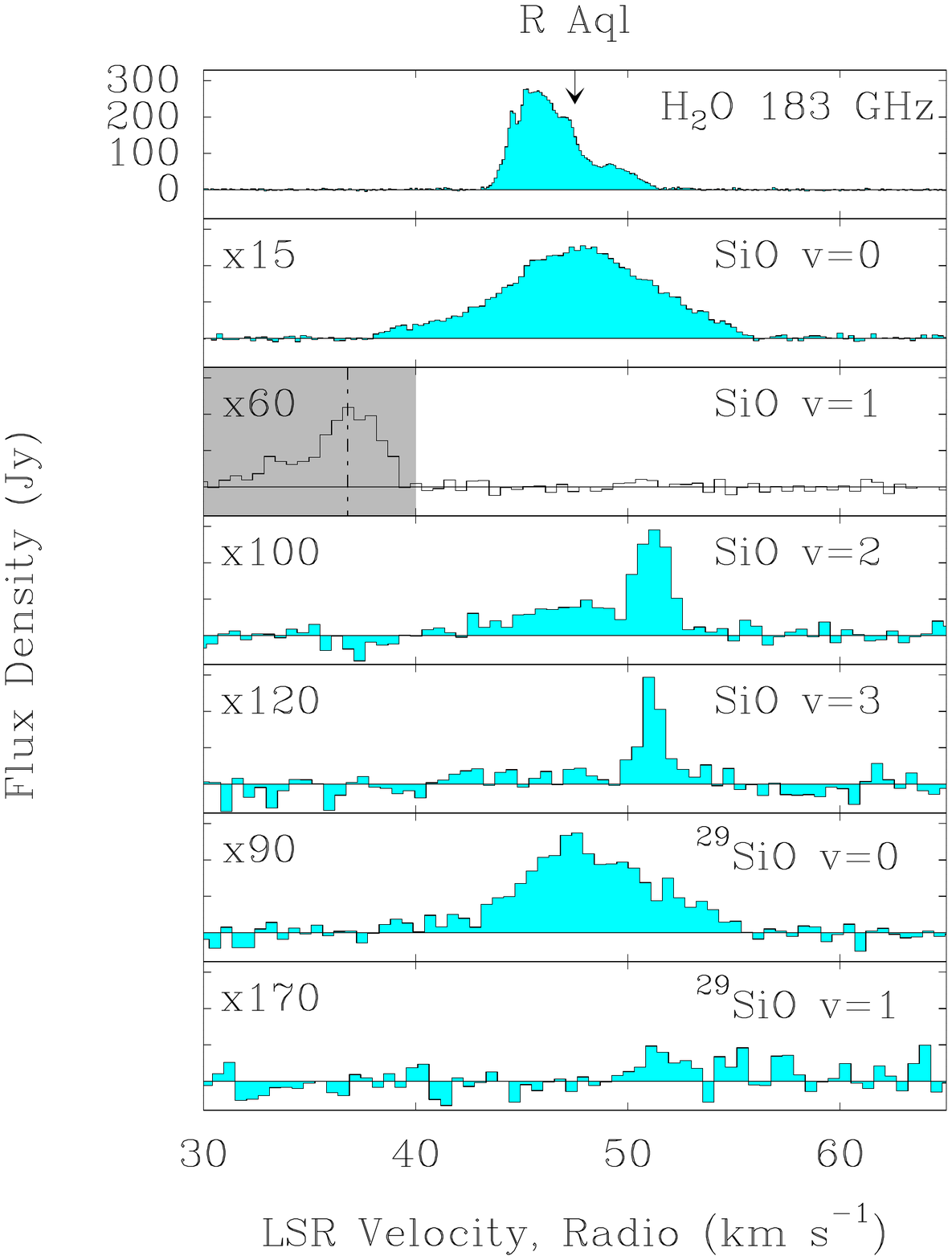}
   \caption{Observations towards R Aql using APEX SEPIA Band 5. See caption of Figure~\ref{spectrum1} for more details.}
              \label{spectrum3}
              \end{figure}

Further analysis was performed in CLASS\footnotemark[2] and
included baseline subtraction of polynomial order 1 for individual scans and spectral smoothing.
The native channel widths of the observations ranged from 0.06 kms$^{-1}$ in the LSB to 0.13 kms$^{-1}$
in the USB. The data were smoothed to velocity resolutions between 0.13 kms$^{-1}$ to 1 kms$^{-1}$.
For the intensity scale a preliminary conversion factor of 34 Jy/K was used across all frequencies. The spectra were 
checked for ghosts (sideband leakage) from the 
other sideband (Sect.~\ref{sideband}).

\footnotetext[2]{http://www.iram.fr/IRAMFR/GILDAS}

The line frequencies of H$_2$O and SiO transitions in the frequency range
of the observations are given in Table~\ref{linelist}. The characteristics of the 
stellar sample, including both AGB and RSG stars, are in Table~\ref{properties}.
The approximate optical stellar phases for the AGB stars at the time of observation are
$\phi$ $\sim$ 0.8 for R Aql and W Hya, and  $\phi$ $\sim$ 0.1 for U Her, determined
using data from the American Association of Variable Star Observers (AAVSO).

\subsection{Sideband Rejection}
\label{sideband}

With the tuning used, it became clear during data reduction that
the water line at 183.310 GHz and the $^{28}$SiO v=1 J=4$-$3 could contaminate each other
due to sideband leakage. \citet{Billade2012} measured the sideband rejection
of the receiver to be $>$10 dB over 90\% of the band, and $>$7 dB over 99\%, with
an 18.5 dB average.
 From our observations 
of R Aql, where the ghost appearing in the LSB from the H$_{2}$O line ($\nu_{obs}$=183.281 GHz) in the USB 
is isolated in velocity, we can
measure a 17.7 dB rejection or leakage at the 2\% level, in good
agreement with the receiver characterization. This value should not be assumed
to hold over all the observations however. In Figures~\ref{spectrum1} to~\ref{spectrum3}, we
have marked the velocity ranges that could be affected by sideband leakage
for each target in grey.

\begin{figure}
   \centering
   \includegraphics[width=9cm]{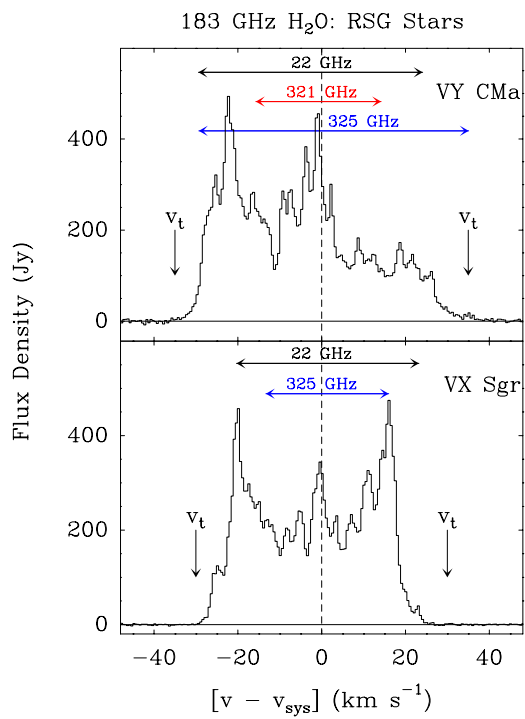}
      \caption{183 GHz H$_{2}$O observations for the RSG stars. Horizontal arrows mark the velocity extent of 22 GHz (black), 321 GHz (red) and 325 GHz (blue) emission measured by \citet{Yates1995}. Note that the 1$\sigma$ rms sensitivity of the observations by \citet{Yates1995} could be up to 14.2 Jy (in 0.33 kms$^{-1}$), so that this should be taken into account when comparing the linewidths. Vertical black arrows mark the terminal velocity, v$_{t}$, estimated for the envelope using low-J CO. For VY CMa,  v$_{t}$ $\sim$ 35 kms$^{-1}$ \citep{Decin2016} and  for VX Sgr v$_{t}$ = 30 kms$^{-1}$ (GA98). }
              \label{rsg_water}
    \end{figure}
    
\begin{figure}
   \centering
   \includegraphics[width=9cm]{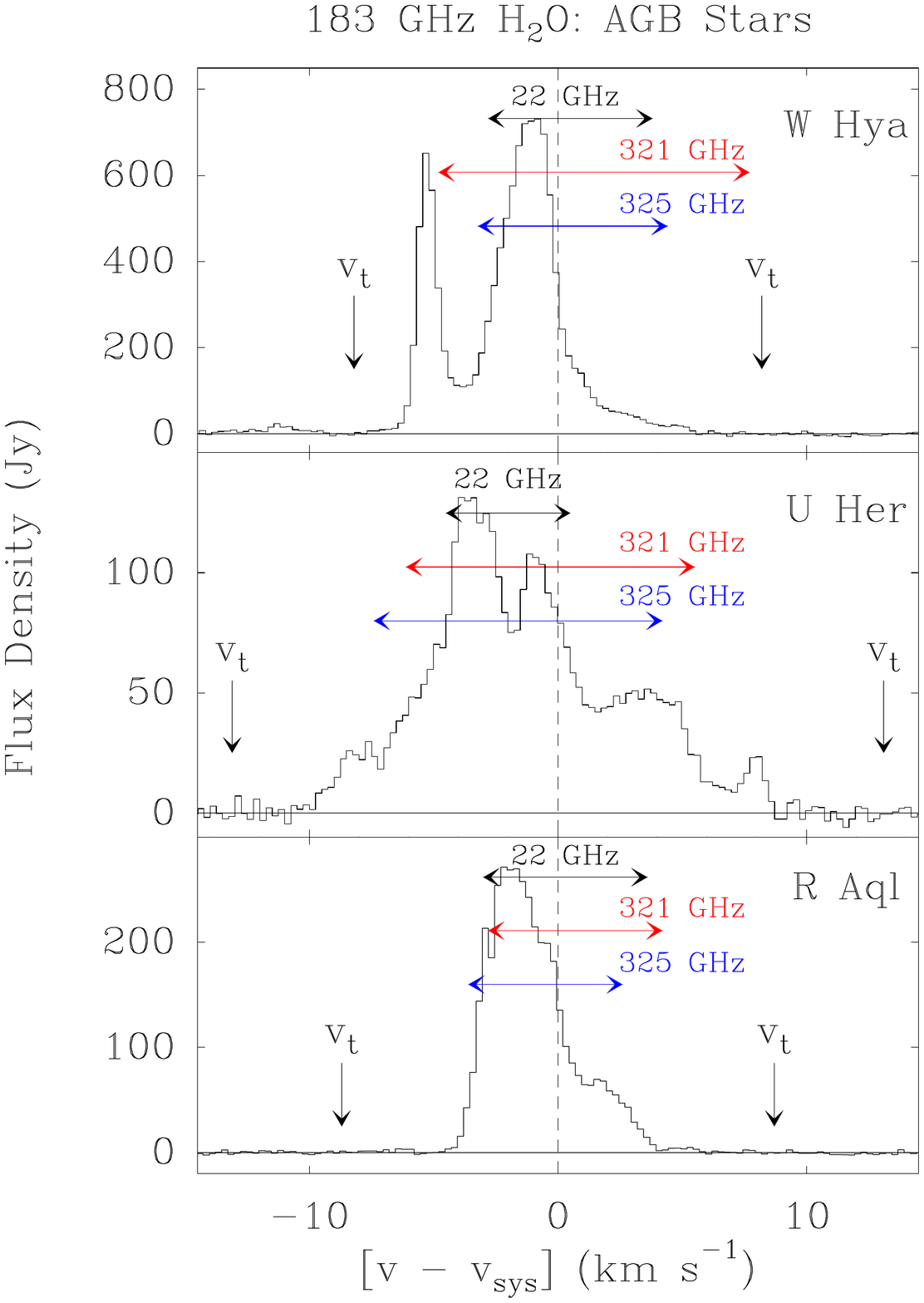}
      \caption{183 GHz H$_{2}$O observations for the AGB stars. Arrows are as for Figure~\ref{rsg_water}. For W Hya v$_{t}$ = 8.2 kms$^{-1}$, for U Her v$_{t}$ = 13.1 kms$^{-1}$
      and for R Aql v$_{t}$ = 8.7 kms$^{-1}$ (GA98).}
              \label{agb_water}
    \end{figure}

\section{H$_{2}$O Emission}

The H$_2$O and SiO spectra are displayed in Figures~\ref{spectrum1} to~\ref{spectrum3}, and the observational results are
summarized in Table~\ref{results}. 
Towards all stars the 183 GHz H$_2$O maser is detected and displays narrow peaks
indicating maser action. 
The velocity extent of the line is similar in each target to the SiO lines that appear to
be dominated by thermal emission, i.e.  the $^{28}$SiO and $^{29}$SiO v=0 lines. The 183 GHz line in VY CMa is a new detection, 
whereas GA98 detected the 183 GHz line towards the other four targets in our sample at more than one epoch. 
Note that
narrow features in the spectra can have FWHM $<$ 1 kms$^{-1}$, e.g. the blue-shifted peak of 183 GHz emission towards W Hya, 
underscoring the importance of observing evolved star masers at high spectral resolution.
We find that the
183 GHz spectra have blue-shifted peaks and peaks near the stellar velocity, but no corresponding red-shifted peak
of similar flux density except in the case of VX Sgr. 

\begin{figure*}
   \includegraphics[width=18.0cm]{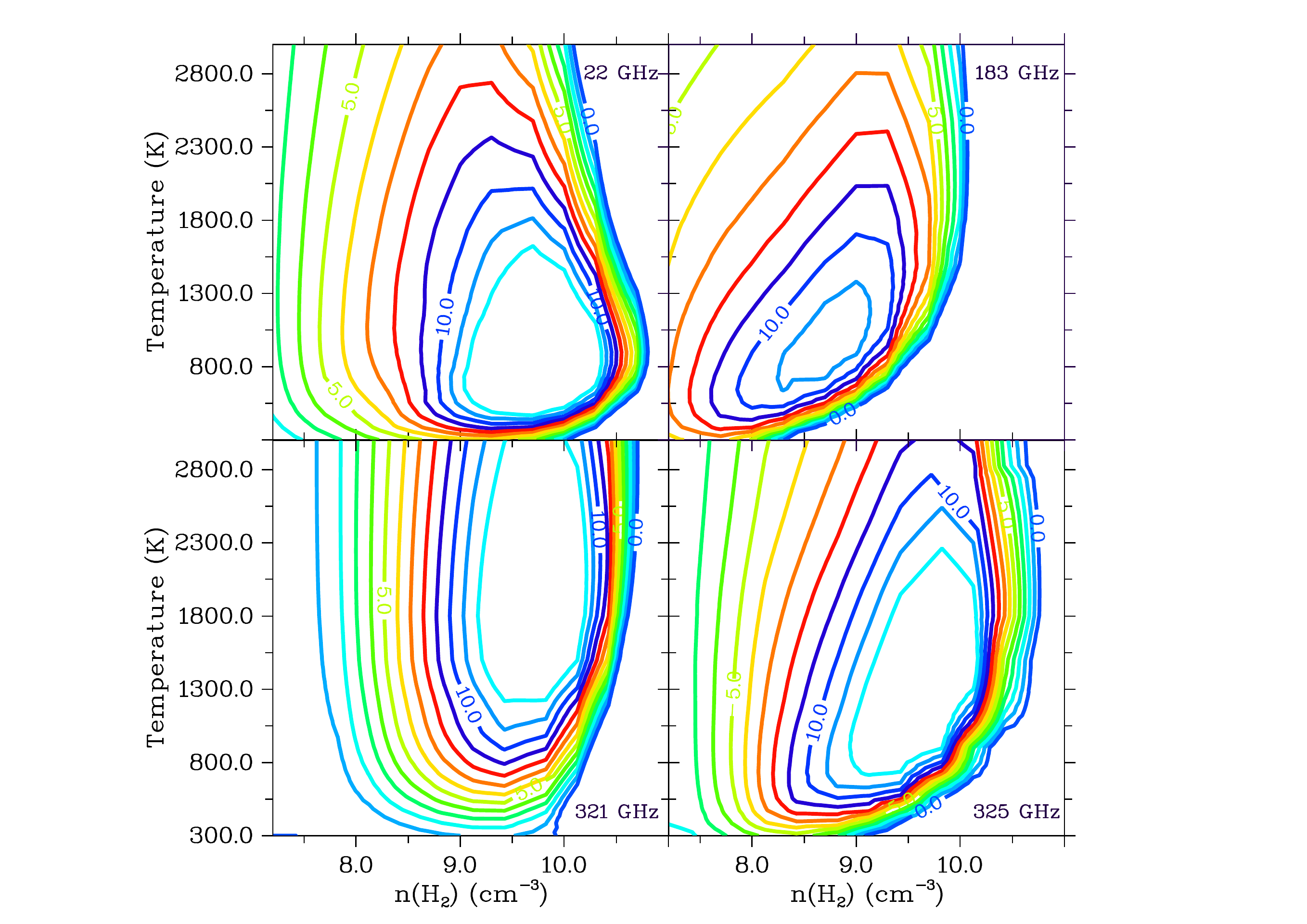}
   \caption{Physical conditions giving rise to circumstellar 183 GHz H$_2$O maser emission from \citet{Gray2016}. Numbers on the contours mark the calculated maser negative optical depth. Strongest 183 GHz H$_2$O maser emission occurs at a lower density
by an order of magnitude than strongest 22 H$_2$O GHz emission.}
              \label{183conditions}%
    \end{figure*}

\begin{figure}
   \centering
  \vspace{-1.5cm}
   \includegraphics[width=9cm]{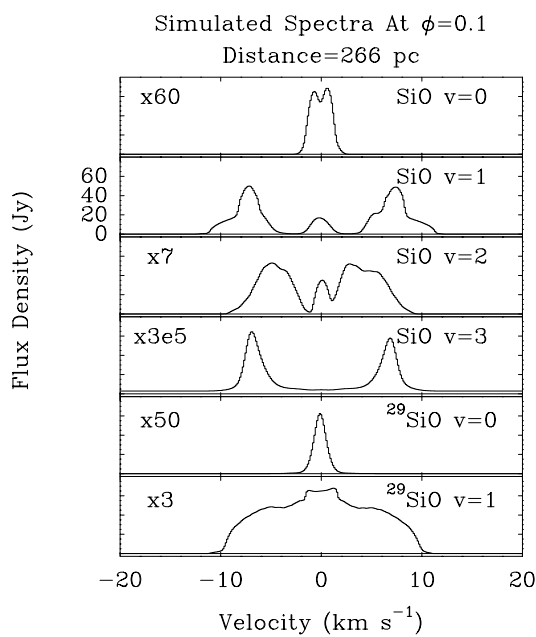}
  \vspace{-1.5cm}
   \caption{Simulated SiO J=4$-$3 spectra from an AGB star at $\phi$=0.1 and a distance of 266 pc for comparison with the U Her observations.}
              \label{fakespectrum1}%
    \end{figure}

\begin{figure}
   \centering
  \vspace{-1.5cm}
   \includegraphics[width=9cm]{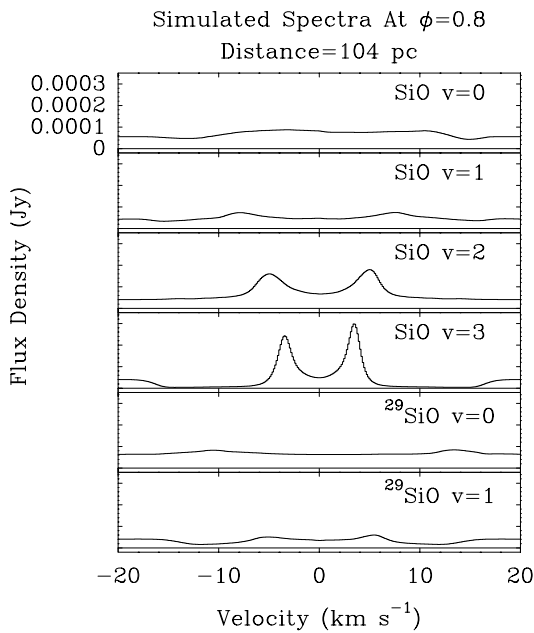}
  \vspace{-1.5cm}
   \caption{Simulated SiO J=4$-$3 spectra from an AGB star at $\phi$=0.8 and a distance of 104 pc for comparison with the W Hya observations. R Aql, at a distance of 422 pc, was also observed at around $\phi$=0.8.}
              \label{fakespectrum2}%
    \end{figure}

\begin{figure*}
   \centering
   {\includegraphics[width=6.5cm]{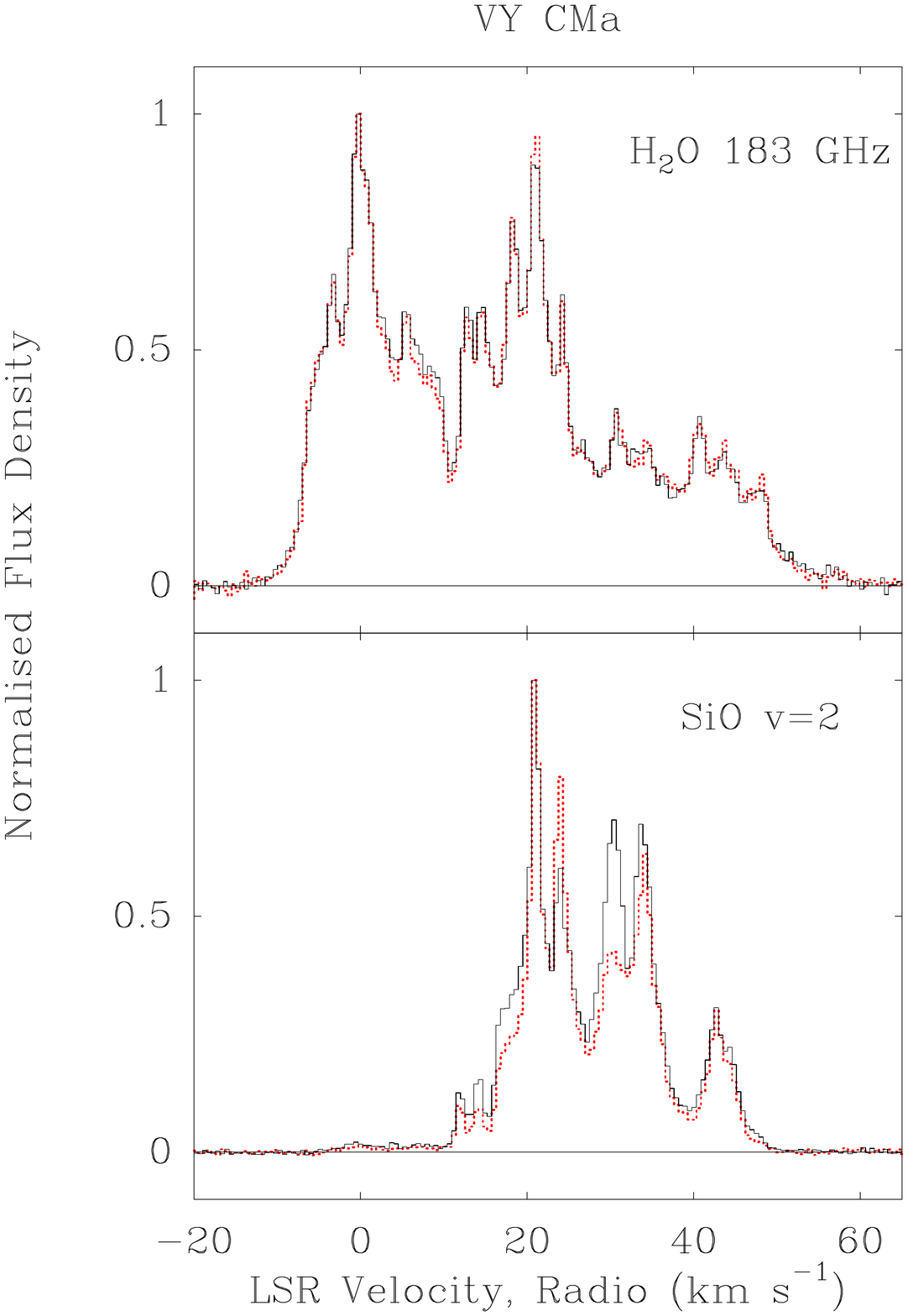}\includegraphics[width=6.5cm]{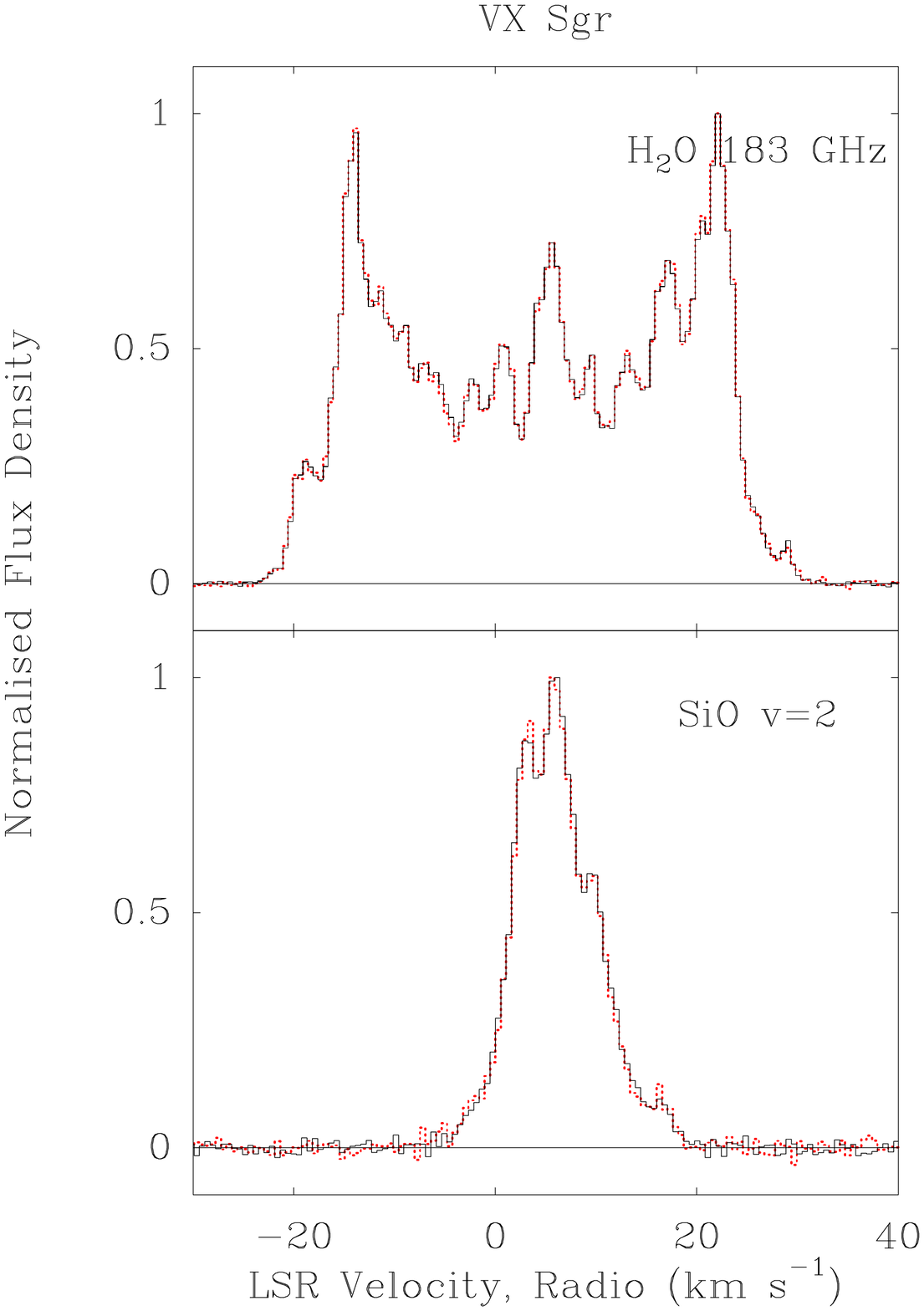}}
   \caption{VY CMa and VX Sgr individual polarizations (solid black and red dotted lines) for the 183 GHz water line and the v=2 J=4$-$3 SiO line. For the purposes of this illustration, each polarization has been normalised to its peak flux density. 
}
              \label{polfigure_rsg}
    \end{figure*}

\subsection{Comparison with previous evolved star observations at 183 GHz}

GA98 made a study of 23 stars using the IRAM 30-m.
They found that, for some objects,  the lineshapes at 183 GHz are dominated by a single peak
near to the stellar velocity. This is indicative of tangential maser amplification close to the star (Group I in GA98). 
Group I object 183 GHz lines
were considerably narrower than corresponding CO (2-1) lines and are associated with objects having mass loss rates $<$ 10$^{-6}$ M$_{\odot}$yr$^{-1}$.
R Aql was found to be part of Group I\footnotemark[3]. 
Group II of GA98 consisted of objects similar in mass-loss rate to Group I, but where the bulk of the emission was
blue-shifted relative to the stellar velocity. However the blue-shifted emission was not close to the terminal velocity of
the envelope and no corresponding red-shifted peak of similar flux density was observed. W Hya and U Her were found to be in Group II.
Finally, Group III of GA98 consisted of objects where the peak emission is shifted to velocities close to the terminal velocity
for the envelope, and there are both red and blue-shifted peaks almost symmetric about the stellar velocity. Usually these
stars have mass-loss rates higher than  10$^{-6}$ M$_{\odot}$yr$^{-1}$. VX Sgr was found to be in Group III.
The observations here are in agreement with these findings, which suggest that 183 GHz maser emission occurs
further from the star as the mass loss rate increases. 
However VY CMa, with the largest mass loss rate, would
not fit the criteria for a Group III object since it does not display a distinct red-shifted peak. The blue-shifted peak
does occur close to the terminal velocity for this object, however, which is at about 35 kms$^{-1}$ \citep{Decin2016}. We speculate
that the terminal velocity for VY CMa is therefore reached relatively close to the star.
Although in principle it would also
possible to draw some conclusions on the variability of the 183 GHz line detected towards VX Sgr,  W Hya, U Her and R Aql, by comparing with the
observations of GA98, in practice this is difficult due to the uncertainty in the fluxscales.

\footnotetext[3]{We note that, using the Hipparcos distance to R Aql, its mass loss rate is $\ge$10$^{-6}$ M$_{\odot}$yr$^{-1}$.}

\subsection{Location of 183 GHz H$_{2}$O emission in the CSE and physical conditions}
      
We do not have simultaneous observation of other water maser transitions towards the targets, needed
to make a detailed line profile comparison between frequencies and constrain line ratios for radiative transfer modelling. 
However, it is possible that the line width of 183 GHz circumstellar water masers is fairly invariant over time.
Comparison of line widths between observations with different sensitivities is difficult to perform, however 
taking into account the higher sensitivity of GA98 we find that the 183 GHz linewidths for VX Sgr, W Hya, 
U Her and R Aql are different by 13\%, 6\%, 8\% and 37\% respectively from those found in this paper.
With this caveat in mind, we go on to compare the line widths at 183 GHz observed here with those from observations
of 22, 321 and 325 GHz water masers by \citet{Yates1995}.

Figures~\ref{rsg_water} and~\ref{agb_water} display the 183 GHz water maser emission we observed towards RSG and AGB stars, with the line widths
for 22, 321 and 325 GHz water masers found by \citet{Yates1995} also indicated. Using linewidth as a proxy for spatial extent in the
CSE, which is valid for emission dominated by regions extending further than about 5 R$_{*}$ (inside this region shock fronts could lead to high velocity
emission that is actually close to the star \citep[e.g.,][]{Herpin1998}), it appears 
that the 183 GHz water maser is at least as extended spatially as these other water masers, and can be more extended.

The discussion of the 183 GHz spectra above supports the predictions of \citet{Humphreys2001}, who used a hydrodynamical pulsating circumstellar envelope model, coupled to an LVG maser
radiative transfer model, to calculate synthetic spectra and maps of 22, 321, 325 and 183  GHz masers for a 1 M$_{\odot}$ AGB star. The 183 GHz 
spectra of AGB stars in this paper are in reasonable agreement with the velocity extent of the synthetic spectrum, which originated from significant 
183 GHz maser emission occurring to relatively far out in the model CSE (20 R$_{*}$).  In addition, \citet{Gray2016} has performed ALI calculations for water maser emission for a grid 
of parameter space appropriate for evolved stars and also finds that 183 GHz masers are pumped over a broad range of
conditions (Figure~\ref{183conditions}), such that the masers can trace the inner CSE and the dust formation zone
and beyond. Followup spatially-resolved observations using ALMA Band 5 are needed to test these predictions. 

The greater difference in velocity extents between 183 GHz and 22 GHz masers for AGB stars versus RGB stars 
  is consistent with the high densities required to pump strong 22 GHz water masers (say 10$^8$ to 10$^{11}$ cm$^{-3}$), 
  which are likely to be confined to a relatively smaller radius in the CSEs of AGB stars than they are in RSG stars. This would be expected considering 
that AGB mass loss rates are lower. On the other hand, 183 GHz masers towards both AGB and RSG could be pumped 
out to relatively far in the CSE since they can operate in significantly lower density and temperature conditions. So the 
velocity extents at 183 GHz towards both types of objects can be relatively broad. We also note that the 22 GHz masers in AGB stars are 
very variable and the observations were not made at the same time as the 183 GHz observations.

\subsection{Relation to OH masers and amplification of the stellar continuum}

In Figure~\ref{spectrum2}, W Hya displays a peak near the stellar velocity and a more blue-shifted, narrow feature at 35.5$\pm$0.1 kms$^{-1}$. 
The FWHM of the blue-shifted feature is $<$ 1 kms$^{-1}$, consistent with it arising from a single maser cloud.
Its velocity is in good agreement with the velocity of the narrow peak in the OH 1667 MHz maser spectrum 
towards this target at 35.6 kms$^{-1}$ \citep{Etoka2001,Vlemmings2003}. 

This suggests that the blue-shifted feature in the 183 GHz H$_{2}$O maser 
spectrum could arise from gas at the same distance in the CSE as that leading to the corresponding OH 1667 MHz emission peak. The emission
from water and OH would not need to arise from the same volumes of gas, if the CSE is clumpy on appropriate spatial scales. Indeed this
is the explanation for how water masers at 22 GHz and mainline OH masers can exist at similar radii despite requiring different collision
partner densities by, say, a factor of 10 or more \citep{Richards1999,Richards2012}. 

However, we note that the situation may be different for the 183 GHz water  maser
emission.
The physical conditions that lead to strong OH maser 1667 MHz  emission in Miras are T$_k$ $\sim$ 100 K, dust temperature T$_{d}$ $\sim$ 350 -- 450 K
and n(H$_{2}$) = 10$^{5}$ cm$^{-3}$ \citep{Nguyen1979,Bujarrabal1980}.  This parameter space for pumping water maser emission at 183 GHz is explored by \citet{Gonzalez1999}, 
at least in terms of gas kinetic temperature and density. \citet{Gonzalez1999} indicates that water maser emission at 183 GHz can result from
similar temperatures and densities as those needed to produce 1667 MHz  OH masers. 
While 183 GHz
water maser emission can also originate from the higher density conditions needed to produce 22 GHz emission  \citep[Figure~\ref{183conditions},][]{Gonzalez1999,Humphreys2001,Gray2016}. Therefore it may
be possible for 183 GHz emission to be arising both from the high and low density gas, as long as sufficient H$_{2}$O abundance remains
in the low density gas and has not been photodissociated by the interstellar radiation field.

VLBA observations by \citet{Vlemmings2003} suggested that the 35.6 km s$^{-1}$  feature in the 1667 MHz maser spectrum is located along the line of sight to the stellar disc 
and could be amplifying the stellar continuum.
We speculate that this is also the case for the 183 GHz water maser feature at the same velocity. If so, we would expect it to vary following the stellar pulsation cycle as the star brightens and dims. 
Single-dish monitoring could investigate if this is the case.

\section{SiO Emission}

The J=4$-$3 SiO lines  are
similar in velocity extent and complexity to those previously observed at multiple frequencies \citep[e.g.,][]{Humphreys1997,Pardo1998,Gray1999}.
The $^{28}$SiO v=0 line appears to be dominated by thermal emission 
for each target, whereas the $^{29}$SiO v=0 looks to be a mixture of both thermal and maser action. Towards VY CMa, the highest mass-loss rate object, this results
in a narrow maser feature superposed on a broad emission plateau. For emission from the v=1,2,3 $^{28}$SiO and the v=1 $^{29}$SiO 
then lineshapes are characteristic of maser emission, and the total velocity extent of the lines is narrower than those from the $v=0$ states.
The only non-detections are for the $^{28}$SiO and $^{29}$SiO v=1 J=4$-$3 lines towards R Aql.
All the SiO lines are weaker than the 183 GHz water emission  observed towards the same object. 

\subsection{Comparison with simulations}

\citet{Gray2009} coupled a Large Velocity Gradient (LVG) $^{28}$SiO maser radiative transfer code to hydrodynamical
models of the CSE by \citet{Ireland2004a,Ireland2004b}, to predict SiO maser spectra and maps at different
epochs of the stellar pulsation cycle. In an update to this work, Gray et al. (in prep) have replaced the LVG code with
Accelerated Lambda Iteration (ALI) SiO radiative transfer. The ALI code performs calculations for $^{28}$SiO, $^{29}$SiO and $^{30}$SiO isotopologues, and includes line overlap between these species.  It can 
handle both thermal and maser emission. The stellar models used with the ALI code are the Cool Opacity-sampling Dynamic EXtended  (CODEX) atmosphere (up to $\sim$5 R$_{*}$) models 
of \citet{Ireland2008,Ireland2011}, specifically the "o54" series. These are meant to describe a star like $o$ Ceti with a period of 330 days.

Although the models cannot be used to make a detailed comparison with these observations, since
we did not observe $o$ Ceti, Figure~\ref{fakespectrum1} shows the results of 
the simulations at $\phi$=0.1 shifted to a distance of 266 pc, i.e. suitable
for comparison with the U Her results. Figure~\ref{fakespectrum2} shows the simulated
spectra for $\phi$=0.8  to compare with W Hya and R Aql.
The synthetic spectra broadly reproduce the complexity and linewidths of the maser lines, although we note that 
the v=3 J=4$-$3 is emission is significantly weaker than observed at $\phi$=0.1 and the higher peak intensity of the
v=2 J=4$-$3 emission than that of the v=1 is also not found in the simulations.  At $\phi$=0.8 the simulated SiO spectra are extremely weak. 
This does not compare well with the observations of W Hya, but for R Aql the observed SiO spectra are indeed relatively weak, with non-detections
of both $^{28}$SiO and $^{29}$SiO v=1 J=4$-$3 emission. We speculate that this indicates R Aql bears more similarity to $o$ Ceti
than does W Hya. 

The simulated thermal lines ($v=0$) in Figures~\ref{fakespectrum1} and~\ref{fakespectrum2} are not similar to the observed ones. This is probably due to the fact the model for 
the inner CSE we are using only goes out to 5 R$_{*}$. Although the gas phase SiO likely depletes
around this radius, due to dust formation, a significant amount of SiO will be present in the gas phase further out. For example \citet{Lucas1992}
imaged SiO v=0 J=2-1 emission towards a number of AGB stars using the Plateau de Bure Interferometer. The diameter of the thermal SiO
emission towards W Hya was 0.9$\pm$0.1 arcseconds, or about 95 AU. Since the stellar radius is W Hya is about 2 AU \citep{Ohnaka2016}, 
there is significant gas phase SiO out to about 45 R$_{*}$.  For M- and S-type stars, \citet{Gonzalez2003} and \citet{Ramstedt2009} 
find e-folding radii for the SiO of 100-1000 AU.

Overall the pulsating models of \citet{Ireland2011} appear to be good
descriptors of the inner CSE, and the conditions leading to SiO emission are
in good agreement with those determined by \citet{Herpin2000}.
Monitoring of the lines and again followup
observations using ALMA will provide further constraints to this work.

\begin{figure*}
   \centering
   {\includegraphics[width=6.5cm]{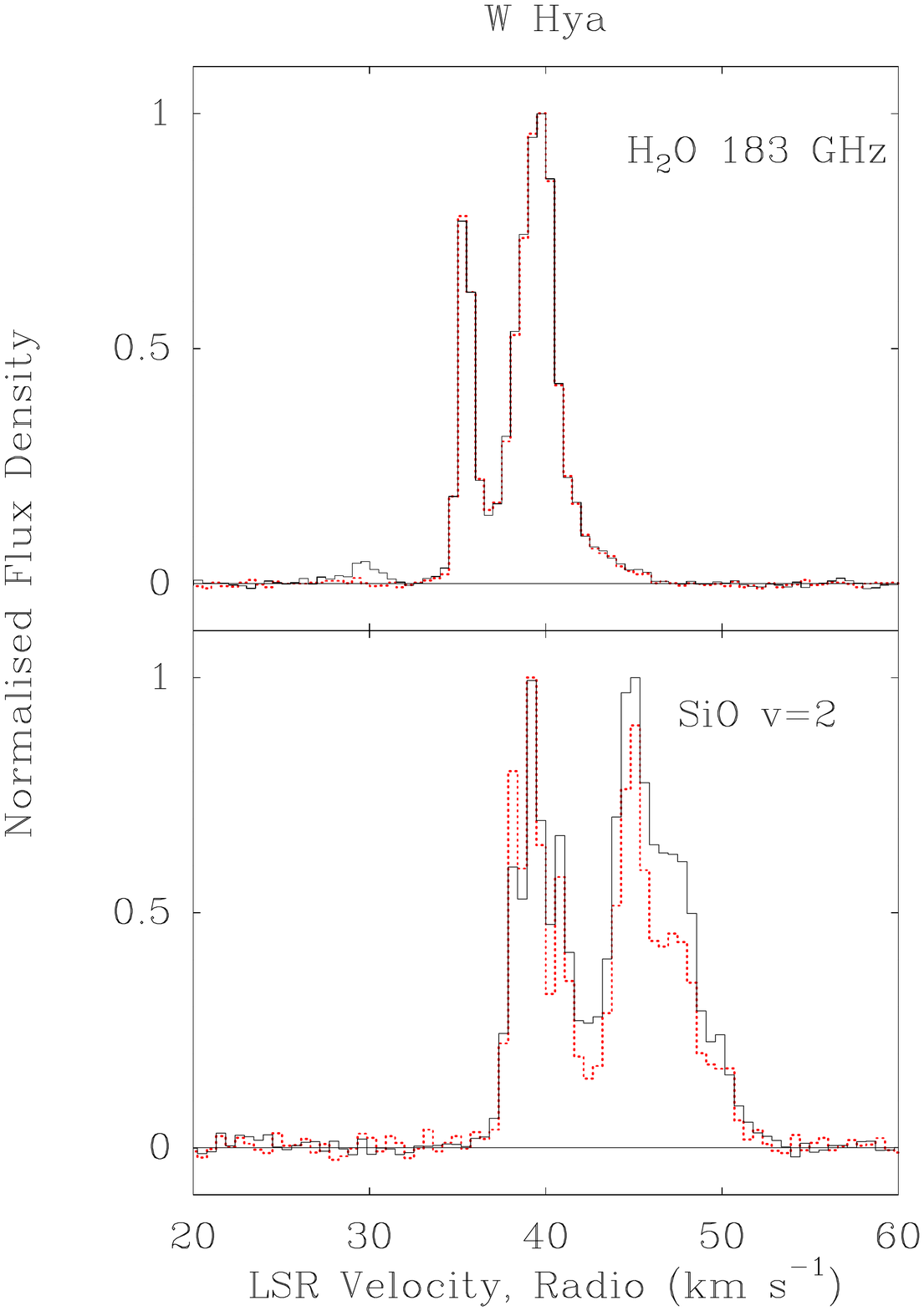}\includegraphics[width=6.5cm]{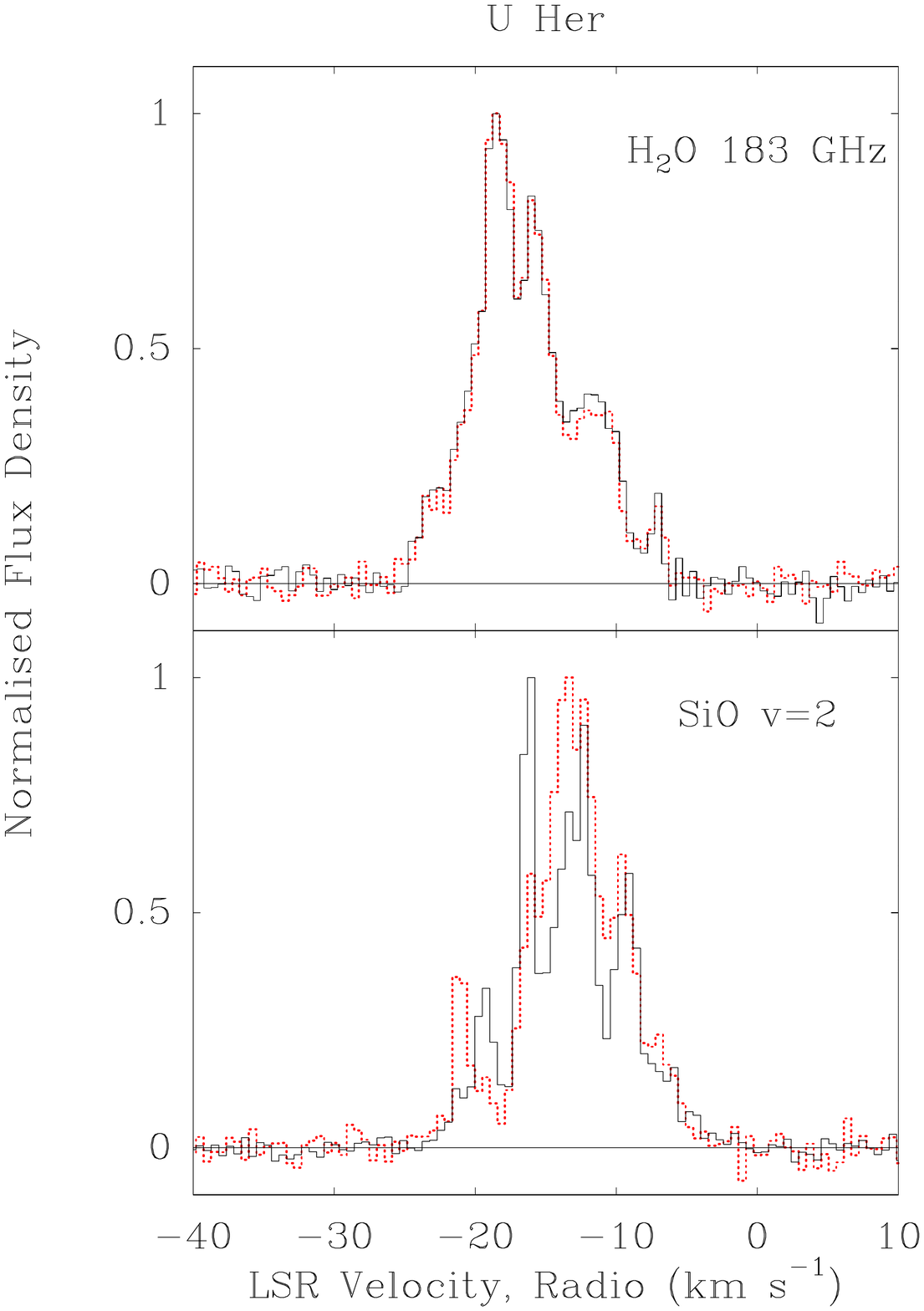}\includegraphics[width=6.5cm]{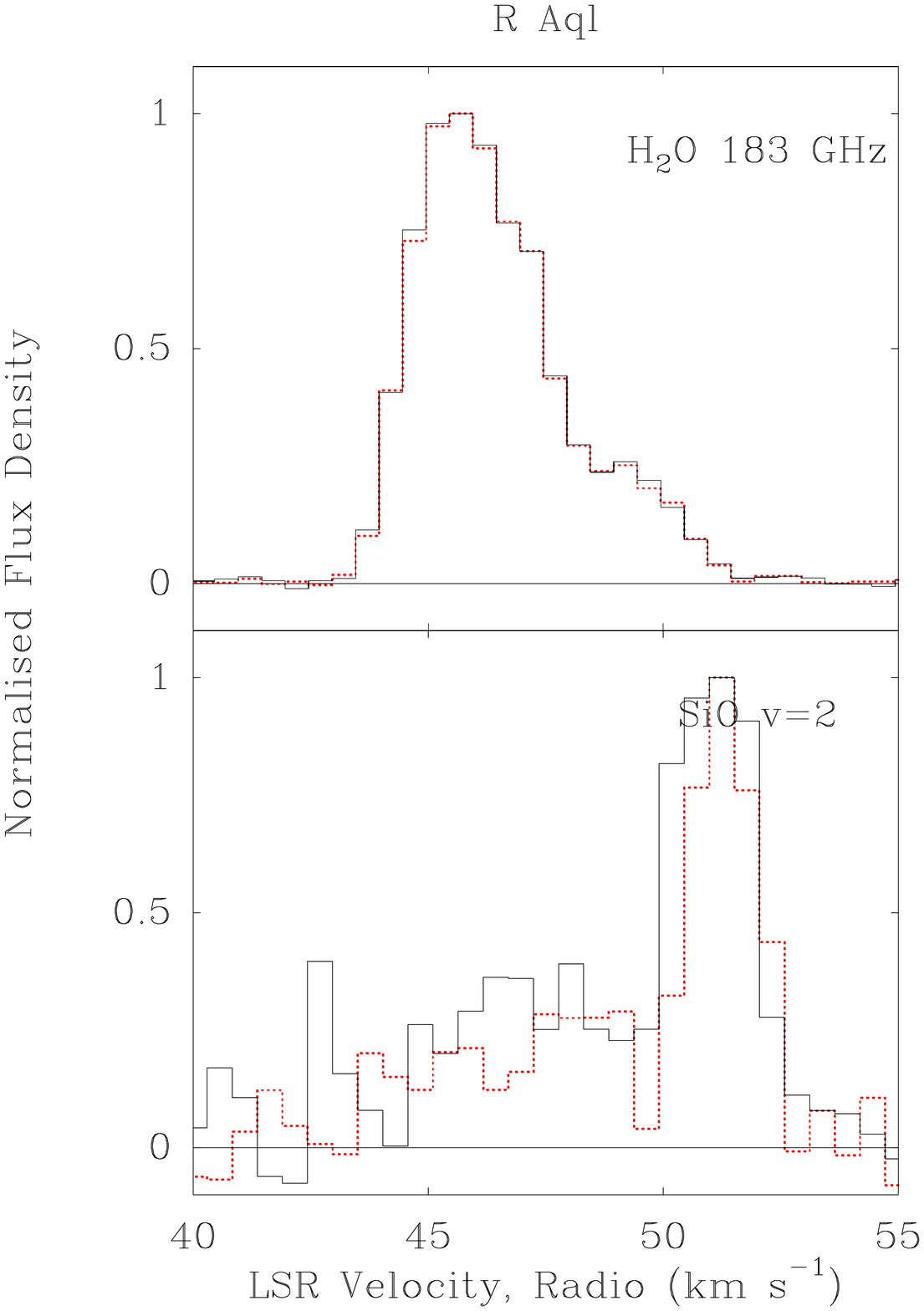}}
   \caption{W Hya, U Her and R Aql individual polarizations (solid black and red dotted lines) for the 183 GHz water line and the v=2 J=4$-$3 SiO line.}
              \label{polfigure_agb}
    \end{figure*}

\section{Line Polarization}

The SEPIA Band 5 receiver records dual linear orthogonal polarizations. This enables the study of 
linear polarization of strong lines. Many masers are strongly linearly polarized 
\citep[e.g.][]{Perez+2013}, making them good probes of magnetic fields. Polarization 
observations of e.g. (sub)millimetre SiO masers have been used to determine the magnetic 
field morphology around evolved stars \citep[e.g.][]{Vlemmings+2011,Shinnaga+2004}, and polarization 
fractions of several tens of percent are regularly measured. Although the observations presented 
here were not aimed at detecting the polarization, an inspection of the two orthogonal polarizations 
independently can already provide a lower limit estimate of the polarization fraction. The difference in 
flux density between the two polarizations depends on the intrinsic source polarization, the maser polarization 
angle (potentially different for different velocity channels) and the amount of signal leakage of one polarization 
into the other. Additionally, as the telescope feed rotates across the sky, with the feed parallactic angle changing 
with respect to the source polarization direction, the signal in the two polarization channels will change as a function 
of parallactic angle. Our observations were restricted to a very narrow range of parallactic angle. We thus cannot determine 
the intrinsic polarization angle, nor the feed leakage, and thus limit ourselves to presenting the potential of the SEPIA receiver 
for maser polarization studies.

In Figures~\ref{polfigure_rsg} and~\ref{polfigure_agb} we show the different polarizations (normalized) for the SiO, v=2 and H$_2$O lines for our sources observed in July 2015. The H$_2$O maser polarization at 183~GHz is significantly lower than that of the SiO, v=2 maser.  Based on an inspection of the spectra, the SiO maser polarization can
reach up to $\sim$30\% and is typically in the range of 5$-$10\%,
while the H$_2$O polarization is not more than a few percent. The low
polarization observed for the H$_2$O masers and the thermal SiO lines
shows that the differential leakage between the two SEPIA polarization
channels is small \citep{Billade2012}. Although the average SiO maser polarization can be
produced under regular, non-saturated, maser assumptions, the observed
maximum linear polarization fraction of some of the individual
features exceeds the maximum fraction allowed under those assumptions \citep{Perez+2013}.

This has previously been noted \citep[e.g.][]{Vlemmings+2011} and requires strong anisotropic pumping of the maser transition \citep{NedoluhaWatson1994}. 
The low polarization fraction of the H$_2$O maser however, fits with the expectation for a non-saturated maser \citep{Perez+2013}. Dedicated observations 
over a large range of parallactic angles will allow us to determine both velocity resolved polarization fraction and linear polarization direction, providing 
constraints on both the circumstellar magnetic field as well as the anisotropic radiation field.

\section{Summary}

We used APEX SEPIA Band 5 Science Verification data to make a simultaneous study of 183 GHz water  and multiple SiO lines
towards five evolved stars. 
Towards each star, we detected 183 GHz water emission with a peak flux density of $>$100 Jy. 
Narrow features in the spectra suggest a maser origin for the lines.
A comparison with velocity widths of water lines at 22, 321 and 325 GHz suggests that the 183 GHz emission is
at least as spatially extended in circumstellar envelopes as these other water maser lines, and may well extend further.
This scenario is supported by radiative transfer models that indicate the 183 GHz maser lines can be pumped out
to zones of relatively low density and kinetic temperature. Towards W Hya, we detect a narrow, blue-shifted feature
at the same velocity as a prominent feature in the 1667 MHz OH maser spectrum that is amplifying the stellar
continuum. We speculate that the blue-shifted 183 GHz water maser feature also amplifies the W Hya stellar continuum. 
For SiO, the lines appear to originate from both thermal and maser action, and sometimes a mixture (e.g. VY CMa
$^{29}$SiO v=0 J=4$-$3). The lines are always weaker than the 183 GHz water maser towards the same object. Typically
the v=1,2 SiO J=4$-$3 lines have the highest peak flux densities, however towards W Hya the v=3 SiO line is stronger.
 From a comparison of the individual polarizations, we find that the SiO maser linear polarization fraction of
several features exceeds the maximum fraction allowed under standard
maser assumptions and requires strong anisotropic pumping of the maser
transition and strongly saturated maser emission.  The low polarization fraction of the H$_2$O maser however, fits with the expectation for a non-saturated maser.
Understanding the locations of the masers in CSE and detailed polarization studies will require high angular resolution observations
using ALMA.

\begin{acknowledgements}
We thank APEX staff for carrying out these observations. We acknowledge with thanks the variable star observations from the AAVSO International Database contributed by observers worldwide and used in this research. We thank staff at ESO and at Onsala Space Observatory, Chalmers University for assistance with the data. WV acknowledges support from ERC consolidator grant 614264.
\end{acknowledgements}

%
%

\bibliographystyle{aa}
\bibliography{sepia_apex}

\begin{sidewaystable*}
           \caption{\label{results} Observational Results}
           \centering
                   {\scriptsize
   \begin{tabular}{ccccccccccccccc} 
   \hline\hline
Star   &\multicolumn{2}{c}{183 GHz}&\multicolumn{2}{c}{$^{28}$SiO v=0}&\multicolumn{2}{c}{$^{28}$SiO v=1}&\multicolumn{2}{c}{$^{28}$SiO v=2}&\multicolumn{2}{c}{$^{28}$SiO v=3}&\multicolumn{2}{c}{$^{29}$SiO v=0} &\multicolumn{2}{c}{$^{29}$SiO v=1}\\
       &\multicolumn{2}{c}{H$_{2}$O}&\multicolumn{2}{c}{J=4$-$3}       &\multicolumn{2}{c}{J=4$-$3}        & \multicolumn{2}{c}{J=4$-$3}        &   \multicolumn{2}{c}{J=4$-$3}     &   \multicolumn{2}{c}{J=4$-$3}     & \multicolumn{2}{c}{J=4$-$3}  \\  
       &    Peak      &  $\Delta$V  &   Peak      &  $\Delta$V       &   Peak      &  $\Delta$V        &     Peak    &  $\Delta$V        &      Peak    &  $\Delta$V        &     Peak    &  $\Delta$V        &   Peak    &  $\Delta$V       \\
       &   (rms)      &  Range  &      (rms)      &  Range       &   (rms)     &  Range            &     (rms)   &  Range            &     (rms)   &  Range             &     (rms)   &  Range             & (rms)   &  Range       \\
       &    (Jy)      & (km s$^{-1}$)&     (Jy)       & (km s$^{-1}$)  &   (Jy)      & (km s$^{-1}$)      &      (Jy)   & (km s$^{-1}$)     &      (Jy)   & (km s$^{-1}$)        &      (Jy)   & (km s$^{-1}$)      &  (Jy)   & (km s$^{-1}$)  \\
    &              &             &                &                &            &                   &             &                   &        &      &   &  & &  \\
      
\hline
    VY CMa &  493.7   & [-13.7, 56.2] &    30.2 & [-22.6, 49.6]  &   96.8   &   [-3.2, 45.1] & 176.9          & [-3.2, 50.7]     &  51.2  &[5.6, 21.2]    & 10.4 & [-13.8, 45.5]   &81.5 & [11.6, 36.3]  \\
                  &   (3.4)    &           &    (0.6) &           &   (0.5)   &       &   (0.5)          &          &  (0.5)      &  &     (0.5)  &  & (0.5) &  \\
\hline
VX Sgr &    474.6   & [-22.8, 32.1] &  8.9       & [-15.6, 30.2]   &  79.2  & [-29.8, 41.3]\tablefootmark{a}    &38.0      & [-7.2, 19.5]  & 8.7        &[0.3, 10.5]  & 3.1  &[-15.0, 30.8] & 37.1&[-2.2, 13.9]   \\
            &      (1.4)   &          &  (0.3)     &          &  (0.3)  &                                         &  (0.3)    &                   & (0.3)       &                  & (0.3)  &  &(0.3) &  \\
\hline
W Hya & 728.6    &[33.3, 46.8]  &    30.8     & [32.6, 48.4]  &  227.8     &  [24.2, 49.1]\tablefootmark{a}   &38.2    &[37.1, 52.0] &  314.1  & [36.9, 50.3] & 19.1  &[34.5, 46.8] & 187.3&[35.9, 47.7]   \\
            &    (2.2)   &                   &     (0.5)    &                      &   (0.4)      &                                                 & (0.4)    &                  &(0.4)      &                    & (0.4) & & (0.4)       &  \\
\hline
U Her  &  130.5  & [-24.5, -6.5]            &    7.1    &[-19.9, -6.2]  &  19.9     &[-19.9, -5.6] &   29.8 & [-22.4, -4.3] &  7.1   & [-16.8, -9.8] &2.8   & [-22.2, -8.9]     &   12.0   & [-16.5, -10.1]  \\ 
           &  (2.2)    &                                &  (0.5)   &                    &  (0.4)    &                   &   (0.4) &                    &  (0.4) &                    & (0.4) &                       &   (0.5)   &  \\
\hline
R Aql  &  270.6 &[43.2, 53.2]  &   16.7    &[38.5, 55.4]  &   $<$0.7   &     ---     &    2.9         &  [42.7, 52.3]  & 2.5   &[49.9, 52.1]   & 3.1   & [43.3, 54.0]&$<$0.6 &---   \\ 
          &  (1.1)  &                     &    (0.2)   &                    &  (0.2)        &             &    (0.2)         &                   &  (0.2) &                     & (0.2)  &                 & (0.2)&  \\
            \hline
 \end{tabular}
}
\tablefoot{RMS values are given for a velocity resolution of 0.5 kms$^{-1}$. Linewidths are measured at the 2$\sigma$ level to approximate to full-width zero maximum.
\tablefoottext{a}{Linewidth appears to contaminated by ghost emission from the other sideband. The true linewidth is narrower.}}
\end{sidewaystable*}

\end{document}